\documentclass[a4paper]{article}
\def\be{\begin{equation}}
\def\eea{\end{eqnarray}}
\def\bea{\begin{eqnarray}}
\def\ee{\end{equation}}
\def\a{\alpha}
\def\b{\beta}
\def\g{\gamma}
\def\s{\sigma}
\usepackage[psamsfonts]{amssymb}
\usepackage{amsmath}
\author  {Y. Naimi$^{a}$ and F. Roshani$^{b,c}$
\\ {\small  $^{a}$Physics Research Center, Islamic Azad University Science and Research Branch,  }
\\ {\small Tehran, Iran}
\\ {\small $^{b}$Department of Physics, Alzahra University, Tehran,
19938-91167, Iran}
\\ {\small $^{c}$ Institute for Studies in Theoretical Physics and
Mathematics (IPM)}
\\ {\small  P.~O.~Box 19395-5531, Tehran, Iran}
\\ {\small E-mail: naimi.y@srbiau.ac.ir and farinaz@ipm.ir}}
\title{Solvable multi-species reaction-diffusion processes, with particle-dependent hopping rates}
\begin{document}
\maketitle
\begin{abstract}
\noindent  By considering the master equation of the totally
asymmetric exclusion process on a one-dimensional lattice and
using two types of boundary conditions (i.e. interactions), two
new families of the multi-species reaction-diffusion processes,
with particle-dependent hopping rates, are investigated. In these
models (i.e. reaction-diffusion and drop-push systems), we have
the case of distinct particles where each particle $A_\alpha$ has
its own intrinsic hopping rate $v_{\alpha}$. They also contain the
parameters that control the annihilation-diffusion rates
(including pair-annihilation and coagulation to the right and
left). We obtain two distinct new models. It is shown that these
models are exactly solvable in the sense of the Bethe anstaz. The
two-particle conditional probabilities and the large-time behavior
of such systems are also calculated.

\noindent{}

\noindent{\bf Keywords}: driven diffusive systems (theory),
quantum integrability (Bethe ansatz), stochastic processes
(theory)
\end{abstract}

\section{Introduction}

Our understanding of non-equilibrium statistical physics is much
behind than that of equilibrium theory. Asymmetric exclusion
processes (ASEP) are paradigmatic models for systems which are far
from equilibrium. Despite their greatly reduced complexity they
capture various fields of physics like the study of shocks
\cite{1,2}, the noisy Burgers equation \cite{3}, polymers in
random media, dynamical models of interface growth \cite{4}, the
traffic models \cite{5}, and the kinetics of biopolymerization
\cite{6}.

 On the one-dimensional lattice, the simplest system is the totally asymmetric exclusion
process (TASEP). In this model, each lattice site is occupied by
at most one particle and all particles can only hop with equal
rate to their right-neighboring site, provided this site is empty;
otherwise the attempted move is rejected. In \cite{7}, TASEP has
been solved by introducing a master equation which describes the
evolution equation of the particles when they are not in
neighboring sites, and a so-called boundary condition, which
specifies the situation in which the probabilities go outside the
physical regions. This happens when some of the particles are in
adjacent sites and the non-physical probability terms appear in
the master equation. The coordinate Bethe ansatz (BA) has been
used to obtain the $N$-particle conditional probabilities of
TASEP.

 The interesting point is that one can find the new solvable
 models (including interactions besides diffusion to right-neighboring
 sites), with other boundary conditions and the same master
 equation. In \cite{8}, a similar technique has been used to solve
 the so-called drop-push model. In this model the particle hops to the next-right site, even if it is occupied.
It can hop by pushing all the neighboring particles to their
next-right sites, with a rate depending on the number of these
particles. Some other generalizations of TASEP
 can be found in \cite{9,10,11}.

  The other interesting point is the study of multi-species systems
 in which several kinds of particles move and interact
 (i.e. reaction-diffusion) on a lattice. In \cite{12}, it has been shown that more-than-one species systems
 are solvable in the sense that the $S$-matrix corresponding to them is
 factorizable into two-particle $S$-matrices. It  has been found there
 that the criterion for this is that the interactions must be such that the $S$-matrix satisfies a kind of spectral Yang-Baxter (SYB) equation.
 The multi-species generalization of the reactions considered in
 \cite{12}, has been studied in \cite{13} and the authors remarked that the SYB reduces to a non-spectral matrix equation. The drop-push
 reaction of  \cite{8}, has been generalized to
 $p$-species in \cite{14}. The most general totally asymmetric
 reaction-diffusion process, including the extended drop-push interactions has been
studied in \cite{15}.

The other family of diffusion system is the processes in which
particles can hop to the next-right and the next-left neighboring
sites. In \cite{7}, the single-species model with only simple
diffusion to the right and left, i.e. partially asymmetric
exclusion process (PASEP), has been studied and in \cite{16}, the
one-species PASEP drop-push model has been described. The
multi-species generalizations of PASEP models in which each
particle hops to the right by rate $D_{R}$ and hops to the left by
rate $D_{L}$, have been studied in \cite{17}. The $D_L=0$ case of
the models studied in \cite{17} leads to the previous mentioned
TASEP models.

In all of the above studies, particles hop randomly in continous
time on the integer lattice $\mathbb{Z}$. A hopping event occurs
independently for each particle with identical rate. The model of
one-dimensional PASEP in which the rates of jump are chosen
randomly at time zero and fixed for the rest of the evolution, was
first introduced by Benjamini et al. \cite{18} who proved the
existence of a critical density $\rho*$ depending only on the
distribution of hopping rates. They also proved for this case the
hydrodynamic limit of an associated zero-rang process (ZRP) which
is obtained from TASEP by identifying particles with the sites of
a new one dimensional lattice and the interparticle distance
(number of empty sites between particles $i$, $i+1$) as occupation
number at site $i$ of that lattice.

This model has received renewed attention because of the
occurrence of a condensation transition analogous to Bose-Einstein
condensation and because of its close relationship with exclusion
processes. An analogous transition also occurs if the rates are
identical for each particle, but dependent on the lattice distance
to the next particle, see  \cite{19} for a recent review (in terms
of the ZRP) and \cite{20,21,22,23,24,25} for the current
developments.

The generalization of TASEP (only for one-species) with constant
hopping rates to the particle-dependent hopping rates (PDHR) has
been studied in \cite{26}. The authors considered the case that a
particle in site $x_i$ has a hopping rate $v_i$ and used the BA to
obtain in a determinant form the exact solution of the master
equation for the conditional probabilities of TASEP with PDHR.
They also derived a determinant expression for the time-integrated
current for a step-function initial state. In the sense of the BA,
the two-species TASEP in which different particles hop with
different rates and fast particles stochastically overtake slow
ones, has been considered by Karimipour in \cite{27}.

In this paper we are going to study the multi-species TASEP with
PDHR. We obtain two new families of multi-species diffusion
processes with PDHR. The first family (i.e. reaction-diffusion
models) has the following reactions
 \bea\label{1}
 A_\alpha\emptyset &\rightarrow
&\emptyset A_\alpha \ \ \ \ {\rm with
 \ rate}\ \ v_{\a},\cr
  A_\alpha A_\beta &\rightarrow &A_\gamma
A_\delta \ \ \ \ {\rm with
 \ rate}\ \ c^{\alpha\beta}_{\gamma\delta}.
\eea and the second family (i.e. drop-push models), describes the
following processes \bea\label{2}
 A_\alpha\emptyset &\rightarrow
&\emptyset A_\alpha \ \ \ \ {\rm with
 \ rate}\ \ v_{\a},\cr
  A_\alpha A_\beta\emptyset &\rightarrow &\emptyset A_\gamma
A_\delta \ \ \ \ {\rm with
 \ rate}\ \ b^{\alpha\beta}_{\gamma\delta},\cr
   &\vdots &
\eea where the dots indicate the other drop-push reactions with
$n$-adjacent particles. In these models, we have the case of
distinct particles where each particle $A_\alpha$ diffuses to the
right with its own intrinsic hopping rate $v_{\alpha}$. We also
consider the annihilation-diffusion (including pair-annihilation
and coagulation to the right and left) extension of these above
models. The annihilation processes are \bea\label{3}
  A_\alpha A_\beta &\rightarrow &\emptyset A_\beta \
\ \ \ {\rm with
 \ rate}\ \ \delta_{\a\b},\cr
  A_\alpha A_\beta &\rightarrow &A_\alpha \emptyset  \ \ \ \ {\rm with
  \ rate} \ \ \gamma_{\alpha\beta},\cr
  A_\alpha A_\beta&\rightarrow &\ \emptyset\ \emptyset  \ \ \ \ {\rm with
  \ rate} \ \ \eta_{\alpha\beta}.
       \eea
We show that the reaction rates of these models (now including the
annihilation processes) must satisfy some specific constraints and
these models are exactly solvable in the sense of the BA provided
a non-spectral matrix equation is satisfied.

The paper is organized as follows. We first introduce two types of
boundary conditions that can be used to obtain two new families of
$p$-species reaction-diffusion processes with PDHR in section 2.
We use the type 1 boundary condition and the type 2 boundary
condition, in terms of two $p^{2}\times p^{2}$ matrices $c$ and
$b$ respectively, to generalize $p$-species reaction-diffusion and
drop-push systems with identical hopping rates into PDHR systems.
In section 3 we consider the annihilation processes extension of
these new models. In section 4 we investigate the BA solution of
the new models and discuss under what conditions, one can use the
BA to find exact solutions. We show that the matrix
$\tilde{c}$($\tilde{b}$) (a version of $c(b)$ that constructs the
$S$-matrix and determines the
 coefficient of BA) must satisfy a non-spectral matrix
equation. Then in section 5 we show that for $p=2$ (for both
models), the specific class of parameters, which corresponds to
the reactions rates together with the annihilation-diffusion
rates, satisfies a non-spectral matrix equation. Finally in
section 6 we study the conditional probabilities of these models
and specially for two-particle systems of section 5, we obtain the
exact expressions and the large-time behavior of such systems.
\section{TASEP-Generalization}
\subsection{Boundary condition}Consider a $p$-species system with particles
$A_1,A_2,\cdots ,A_p$. The basic objects we are interested in are
the probabilities $P_{\alpha _1\cdots\alpha _N}(x_1,\cdots
,x_N;t)$ for finding at time $t$ the particle of type $\alpha _1$
at site $x_1$, particle of type $\alpha _2$ at site $x_2$, etc. We
take the physical region of coordinates as $x_1<x_2<...<x_N$. The
master equation for a totally asymmetric exclusion process, with
particle-dependent hopping rates, is
 \bea\label{4}
 {\partial\over{\partial t}}P_{\alpha _1\cdots\alpha_N} (x_1,\cdots ,x_N;t)
 &=&\sum_{i=1}^Nv_{\alpha_i} P_{\alpha
_1\cdots\alpha_N}(x_1,\cdots,x_{i-1},x_i-1,x_{i+1},\cdots,x_N;t)\cr
  &&-(\sum_{i=1}^Nv_{\alpha_i})P_{\alpha _1\cdots\alpha_N} (x_1,\cdots ,x_N;t)
   \eea
This equation describes a collection of $N$-particle that the
${\alpha_i}$-th particle drifts to the next-right site by rate
$v_{\alpha_i}$ where $v_{\a_{i}}$s are finite non-zero real
numbers. This master equation is only valid for $x_i<x_{i+1}-1$.
For $x_i=x_{i+1}-1$, there will be some terms with $x_i=x_{i+1}$
in the right-hand side of (\ref{4}), which are out of the physical
region. But one can assume that (\ref{4}) is valid for all the
physical regions $x_i<x_{i+1}$ by imposing certain boundary
conditions for $x_i=x_{i+1}$. Different boundary conditions
introduce different interactions for particles. Following the same
argument given in \cite{17}, it can be easily seen that the master
equation (\ref{4}) leads to the following relation for
two-particle probabilities
 \bea\nonumber
{\partial \over{\partial
 t}}\sum_{x_2}\sum_{x_1<x_2}P_{\a_1\a_2}(x_1,x_2;t)&=&
 \sum_{x}v_{\alpha_2}P_{\a_1\a_2}(x,x;t)
 -\sum_{x} v_{\alpha_1}P_{\a_1\a_2}(x,x+1;t)
   \eea
\be\label{5} \ee
   This equation leads us to take $P_{\a_1\a_2}(x,x;t)$ as the following two
   types of boundary conditions
   \be\label{6}
v_{\alpha_2} P_{\a_1\a_2}(x,x) =\sum_\b c^{\b_1\b_2}_{\a_1\a_2}
 P_{\b_1\b_2}(x,x+1)\ \ \ \ \ \ {\rm type\  1\ \ }
 \end{equation}
  \be\label{7}
v_{\alpha_2} P_{\a_1\a_2}(x,x) =\sum_\b b^{\b_1\b_2}_{\a_1\a_2}
 P_{\b_1\b_2}(x-1,x)\ \ \ \ \ \ {\rm type\  2\ \ }
 \end{equation}
 $\b$ stands for $(\b_1\b_2)$ and $b$ and $c$ are $p^2\times p^2$ matrices which determine the
 interactions. In the probabilities appear in (\ref{6}) and (\ref{7}), we have
 suppressed all the other coordinates and the time $t$ for simplicity.
 We consider the processes in which the number of particles is
 constant in time, in other words we exclude the creation and
 annihilation processes (in fact in this step). Therefor if we sum (\ref{5}) over
 $\alpha_1$ and $\alpha_2$, the left-hand side becomes zero and
 results in

  \be\label{8}
 -\sum_x\sum_\a v_{\a_1}P_{\a_1\a_2}(x,x+1)+
 \sum_x\sum_\b\left(\sum_\a
 c^{\b_1\b_2}_{\a_1\a_2}\right)
 P_{\b_1\b_2}(x,x+1)=0.
 \end{equation}
in which (\ref{6}) has been used. Clearly (\ref{8}) gives
 \be\label{9}
 \sum_\a c^{\b_1\b_2}_{\a_1\a_2}=v_{\b_1} \ \ \ \ {\rm constraint\ for\ type\ 1.\ \ }
 \end{equation}
 In the same way, if we use (\ref{7}), the sum over the elements of each column of matrix $b$ results in
\be\label{10}
 \sum_\a b^{\b_1\b_2}_{\a_1\a_2}=v_{\b_1} \ \ \ \ {\rm constraint\ for\ type\ 2. \ \ }
 \end{equation}
Remember that the constraints (\ref{9}) and (\ref{10}) are the
consequence of the conservation of probabilities so they can be
interpreted as the probability conservation equations.

 \subsection{$P$-species reaction-diffusion systems with PDHR}
 Following the same steps as \cite{17}, we first consider ${\dot
P}_{\a_1\a_2}(x,x+1)$. Using (\ref{4}) and the type 1 boundary
condition (\ref{6}), it is found \bea\label{11}
 {\dot P}_{\a_1\a_2}(x,x+1)&=&
 v_{\a_1}P_{\a_1\a_2}(x-1,x+1)+
 \sum_{\b\neq\a} c^{\b_1\b_2}_{\a_1\a_2}P_{\b_1\b_2}(x,x+1)\cr&-&
 v_{\a_2}P_{\a_1\a_2}(x,x+1)-\sum_{\b\neq
\a} c^{\a_1\a_2}_{\b_1\b_2}P_{\a_1\a_2}(x,x+1).
 \eea
in which we have used (\ref{9}). It can be written as (in fact the
diagonal elements are not reaction rates) \be\label{12}
 c^{\a_1\a_2}_{\a_1\a_2}= v_{\a_1}-  \sum_{\b\neq \a}
 c^{\a_1\a_2}_{\b_1\b_2},
 \end{equation}
It is seen that the evolution equation (\ref{11}) describes the
following two-particle interactions

 \bea\label{13}
 A_\alpha\emptyset &\rightarrow
&\emptyset A_\alpha \ \ \ \ {\rm with
 \ rate}\ \ v_{\a},\cr
  A_\alpha A_\beta &\rightarrow &A_\gamma
A_\delta \ \ \ \ {\rm with
 \ rate}\ \ c^{\alpha\beta}_{\gamma\delta}.
\eea It is simple to show that our formalism is consistent for
more-than-two particle systems. In this model, in addition to
interactions, we have the case of distinct particles where each
particle $A_\a$ has its own intrinsic hopping rate $v_{\a}$. It is
more general than TASEP model has been studied in \cite{13}, in
which all particles have the equal rate.
\subsection{$P$-species drop-push systems with PDHR}
In the same way, we first consider ${\dot P}_{\a_1\a_2}(x,x+1)$ by
using (\ref{4}) and the type 2 boundary condition (\ref{7}). The
result is \bea\label{14}
 {\dot P}_{\a_1\a_2}(x,x+1)&=&
 v_{\a_1}P_{\a_1\a_2}(x-1,x+1)+
 \sum_{\b} b^{\b_1\b_2}_{\a_1\a_2}P_{\b_1\b_2}(x-1,x)\cr&-&
 v_{\a_2}P_{\a_1\a_2}(x,x+1)-\sum_{\b} b^{\a_1\a_2}_{\b_1\b_2}P_{\a_1\a_2}(x,x+1).
 \eea
 in the above equation we have directly used constraint
(\ref{10}) as
 \be\label{15}
 \sum _{\b}b^{\a_1\a_2}_{\b_1\b_2}= v_{\a_1}.
 \end{equation}
 so in agreement with (\ref{14}), the allowed processes are

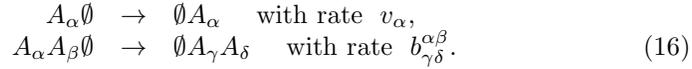
\bea\label{16}
 A_\alpha\emptyset &\rightarrow
&\emptyset A_\alpha \ \ \ \ {\rm with
 \ rate}\ \ v_{\a},\cr
  A_\alpha A_\beta\emptyset &\rightarrow &\emptyset A_\gamma
A_\delta \ \ \ \ {\rm with
 \ rate}\ \ b^{\alpha\beta}_{\gamma\delta}.
\eea  In this model any particle can hop to the right site, with
rate depends on the type of the particle, if that site is empty.
If the right site is occupied, the left particle can still hop to
that site by pushing the right one, but in the mean time there is
a probability that the types of the particles change. It is
important to note that all elements of $b$ (including the diagonal
elements) should be nonnegative (since they are rates). We will
consider more-than-two particle systems when we introduce the type
2 model.

\section{Annihilation-diffusion processes}
The annihilation process is a diffusion-limited reaction-diffusion
process. In this process, the particles annihilate pairwise or
coagulate to the right and left whenever they meet each other. Now
we add the annihilation-diffusion to the previous reactions of
both models. Note that the annihilations appear only in the sink
terms of the evolution equation, as if we consider the initial
state with $n$ particles, no annihilation processes can lead to a
$n$-particle state at any later time. So we do not have the
conservation of probabilities (particles) and one can enter a sink
term, $\lambda_{\b_1\b_2}$, into the conservation equations
(constraints) as

\be\label{17}
 \sum_\a c^{\b_1\b_2}_{\a_1\a_2}=v_{\b_1}-\lambda_{\b_1\b_2} \ \ \ \ {\rm \ \ }
 \end{equation}

\be\label{18}
 \sum_\a b^{\b_1\b_2}_{\a_1\a_2}=v_{\b_1}-\lambda_{\b_1\b_2} \ \ \ \ {\rm \ \ }
 \end{equation}
and if we use these modified constraints in the calculation of
$\dot{P}_{\a_1\a_2}(x,x+1)$, we find the same equation as
(\ref{11}) or (\ref{14}) except extra term
$\lambda_{\a_1\a_2}P_{\a_1\a_2}(x,x+1)$ which is added to the sink
terms. So $\lambda_{\a_1\a_2}$ is the sum of the rates of all
annihilation processes with initial state $(\a_1,\a_2)$ and
therefor it is a positive quantity. These processes are
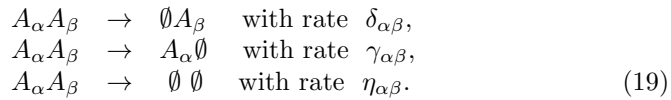
\bea\label{19}
  A_\alpha A_\beta &\rightarrow &\emptyset A_\beta \
\ \ \ {\rm with
 \ rate}\ \ \delta_{\a\b},\cr
  A_\alpha A_\beta &\rightarrow &A_\alpha \emptyset  \ \ \ \ {\rm with
  \ rate} \ \ \gamma_{\alpha\beta},\cr
  A_\alpha A_\beta&\rightarrow &\ \emptyset\ \emptyset  \ \ \ \ {\rm with
  \ rate} \ \ \eta_{\alpha\beta}.
       \eea
and the relation between the above rates is \be\label{20}
\lambda_{\a\b}=\delta_{\a\b}+\gamma_{\a\b}+\eta_{\a\b}.
\end{equation}

One should note that the $p$-species reaction-diffusion and
drop-push systems, including the annihilations, have three kinds
of processes. The first one is the pure diffusion with PDHR which
occurs when a particle is adjacent to a hole, the second and third
ones are the reactions and the annihilations respectively, which
occur when two particles are adjacent to each other. The master
equation for the diffusion with PDHR is (\ref{4}) and for this
case the master equation does not contain anything about both the
reactions and the annihilations. The effect of reactions has been
coded in the boundary condition which is in fact the part of the
master equation. Also the effect of annihilations has been coded
in the modified constraint (\ref{17}) or (\ref{18}).

\section{The Bethe ansatz solution}
\subsection{Solvability criteria}To solve the master equation (\ref{4}) with two types
of boundary conditions, we consider the modified Bethe ansatz of
the form \cite{27}
 \begin{equation}\label{21}
P_{\alpha_1,\cdots ,\alpha _N}({\mathbf x};t)=( \prod_{i=1}^N
v_{\a_i}^{x_i}e^{-v_{\a_{i}}t} )e^{-E_{N}t} \psi_{\alpha_1,\cdots
,\alpha _N}({\mathbf x};t),
\end{equation}
with $\Psi$ as Bethe wave function
\begin{equation}\label{22}
\Psi({\mathbf x})=\sum_\sigma {\mathbf{A}}_\sigma e^{i\sigma
({\mathbf{p}}).{\mathbf{x}}}.
\end{equation}
$\Psi$ is a tensor of rank $N$ with components $\psi
_{\alpha_1\cdots\alpha _N}({\mathbf x})$, where the sum is taken
over all permutations $\sigma$ of $(1,2,3,...,N)$. Inserting
(\ref{21}) in (\ref{4}) therefor the above eigenfunctions
correspond to the eigenvalues as follows
\begin{equation}\label{23}
 E_N =-\sum_{j=1}^Ne^{-ip_j}
\end{equation}
The next step is to determine the coefficients
${\mathbf{A}}_\sigma$. Inserting (\ref{21}) in the type $1$
boundary condition (\ref{6}) gives
\begin{equation}\label{24}
v_{\a_2}e^{-(E_2+v_{\a_1}+v_{\a_2})t}v_{\a_1}^{x}v_{\a_2}^{x}\psi_{\a_1,\a_2}(x,x)=\sum_{\beta}c_{\a_1\a_2}^{\b_1\b_2}e^{-(E_2+v_{\b_1}+v_{\b_2})t}v_{\b_1}^{x}v_{\b_2}^{x+1}\psi_{\b_1,\b_2}(x,x+1)
\end{equation}
relation (\ref{24}) is the boundary condition for $\Psi$ but it
does not has the form of boundary condition (\ref{6}) (because of
the time dependency and the elements of power $x$). To build a
modified form of (\ref{24}) (similar to (\ref{6})), we should omit
the time dependency and the elements of power $x$ from both-hand
sides of this equation. After finding a modified form, we can put
(\ref{22}) in it to obtain the coefficients ${\mathbf{A}}_\sigma$.
Obviously to obtain a modified desired form one arrives at
\begin{equation}\label{25}
v_{\a_1}+v_{\a_2}=v_{\b_1}+v_{\b_2}, \end{equation}
\begin{equation}\label{26}
v_{\a_1}v_{\a_2}=v_{\b_1}v_{\b_2}. \end{equation} one can write
these equations as
\begin{equation}\label{27}
v_{\a_1}+v_{\a_2}=v_{\b_1}+v_{\b_2}, \end{equation}
\begin{equation}\label{28}
v_{\a_1}^2+v_{\a_2}^2=v_{\b_1}^2+v_{\b_2}^2. \end{equation} These
equations have a physical interpretation in terms of an
\textit{elastic collision}. In the elastic collision of equal
masses (supposing that $v_{\b_i}$s are the velocities before
collision and $v_{\a_i}$s are the velocities after collision) the
conservation of momentum and kinetic energy are the same as
(\ref{27}) and (\ref{28}) respectively. Solving these simultaneous
equations we get
 \bea\label{29}
&\{v_{\a_1}=v_{\b_1} ,v_{\a_2}=v_{\b_2} \}\ \ \ \ {\rm and} \ \ \
\ \{v_{\a_1}=v_{\b_2} ,v_{\a_2}=v_{\b_1} \} \eea Two above
solutions imply that only two states for $\vec{\b}$ are
$\vec{\beta}=(\a_1,\a_2)$ and $\vec{\beta}=(\a_2,\a_1)$. These
states introduce a system with PDHR and the
\textit{exchange-reaction} processes as we will show later on. So
the matrix $c$ has non-zero elements provided
$\vec{\beta}=(\a_1,\a_2)$ and $\vec{\beta}=(\a_2,\a_1)$ are our
states and therefore (\ref{24}) reduces to \bea\label{30}
\psi_{\a_1\a_2}(x,x)&=&
c_{\a_1\a_2}^{\a_1\a_2}\psi_{\a_1\a_2}(x,x+1)+(\frac{v_{\a_1}}{v_{\a_2}})c_{\a_1\a_2}^{\a_2\a_1}\psi_{\a_2\a_1}(x,x+1)
\cr &=&\sum_{\beta} \ \ \ \ \ \alpha+\beta>\gamma
\tilde{c}_{\a_1\a_2}^{\b_1\b_2}\psi_{\b_1\b_2}(x,x+1) \eea
where\begin{equation}\label{31}
\tilde{c}_{\a_1\a_2}^{\b_1\b_2}=\left\{\begin{array}{lcr}c_{\a_1\a_2}^{\a_1\a_2}&
\ \ \ for \ \ \ \vec{\beta} = \vec{\alpha}
\\(\frac{v_{\a_1}}{v_{\a_2}})c_{\a_1\a_2}^{\a_2\a_1}& \ \ \ for \ \ \ \vec{\beta}= (\a_2,\a_1).\end{array}\right.
\end{equation}in the same way, for the type $2$ boundary condition
(\ref{7}), one arrives at\begin{equation}\label{32}
\tilde{b}_{\a_1\a_2}^{\b_1\b_2}=\left\{\begin{array}{lcr}(\frac{1}{v_{\a_1}v_{\a_2}})b_{\a_1\a_2}^{\a_1\a_2}&
\ \ \  for \ \ \ \vec{\beta} = \vec{\alpha}
\\(\frac{1}{v_{\a_2}^2})b_{\a_1\a_2}^{\a_2\a_1}& \ \ \ for \ \ \ \vec{\beta}= (\a_2,\a_1).\end{array}\right.\end{equation}

In the $p$-species case, $c(\tilde{c})$ has $p!(p-1)$ non-diagonal
non-zero elements that these elements are the exchange-rates of
the $p$-species particles but $b(\tilde{b})$ has $p!(p-1)+p^2$
non-zero elements ($p^2$ diagonal and $p!(p-1)$ non-diagonal) that
describe the rates of the drop-push interactions. One should note
that the annihilation rates are not the original elements of the
matrix $c(b)$ and they enter the matrix when we impose the
constraint (\ref{17})(constraint (\ref{18})) to this matrix. Now
we can write (\ref{24}) in the compact notation as \be\label{33}
  \Psi(\cdots ,x_k=x,x_{k+1}=x,\cdots)=\tilde{c}_{k,k+1}
  \Psi(\cdots ,x_k=x,x_{k+1}=x+1,\cdots ).
\end{equation} where $\tilde{c}_{k,k+1}$
\be\label{34} \tilde{c}_{k,k+1}=1\otimes\cdots\otimes
 1\otimes\underbrace{\tilde{c}}_{k,k+1}\otimes 1 \otimes\cdots \otimes 1.
\end{equation}
Note that however (\ref{29}) leads to the modified desired form of
(\ref{24}) but in itself implies some loss of generality on the
elements of the matrices $b$ and $c$. Now the coefficients
${\mathbf{A}}_\sigma$ can be determined by
 putting (\ref{22}) in (\ref{33}), which gives \be\label{35}
 [1-e^{i\sigma (p_{k+1})}\tilde{c}_{k,k+1}]
 {\mathbf A}_\sigma +
 [1-e^{i\sigma (p_{k})}\tilde{c}_{k,k+1}]
 {\mathbf A}_{\sigma\sigma_k}=0.
 \end{equation}from this one obtains
 \be\label{36}
 {\mathbf A}_{\sigma\sigma_k}=S_{k,k+1}^{(1)}(\s (p_k),\s (p_{k+1}))
 {\mathbf A}_{\sigma},
 \end{equation}
where the matrix $S^{(1)}$ is defined through
 \be\label{37}
 S^{(1)}(z_1,z_2)=-(1-z_1\tilde{c})^{-1}(1-z_2\tilde{c}).
 \end{equation}
and the definition of $S^{(1)}_{k,k+1}$ is similar to that of
$\tilde{c}_{k,k+1}$ in (\ref{34}). We have also used
$z_k=e^{ip_k}$.
 The same procedure for the type 2 boundary condition (\ref{7})
 results in
\be\label{38}
 S^{(2)}(z_1,z_2)=-(1-z_{2}^{-1}\tilde{b})^{-1}(1-z_{1}^{-1}\tilde{b}).
 \end{equation}
 Equation (\ref{36}) allows one to compute all the
${\mathbf A}_{\sigma}$'s in terms of ${\mathbf A}_{1}$(which is
set to unit). As the generators of the permutation group satisfy
$\s_k\s_{k+1}\s_k=\s_{k+1}\s_k\s_{k+1}$, so one also needs
\be\label{39}
 {\mathbf A}_{\s_k\s_{k+1}\s_k}={\mathbf A}_{\s_{k+1}\s_k\s_{k+1}}.
 \end{equation}
This in terms of $S$-matrices becomes \be\label{40}
 S_{12}(z_2,z_3)S_{23}(z_1,z_3)S_{12}(z_1,z_2)=
 S_{23}(z_1,z_2)S_{12}(z_1,z_3)S_{23}(z_2,z_3).
 \end{equation}
Writing the $S$-matrix as the product of the permutation matrix
$\Pi$ and an $R$-matrix,
 \be\label{41}
 S_{k,k+1}=:\Pi_{k,k+1}R_{k,k+1},
 \end{equation}
Equation (\ref{40}) is transformed to
 \be\label{42}
 R_{23}(z_2,z_3)R_{13}(z_1,z_3)R_{12}(z_1,z_2)=
 R_{12}(z_1,z_2)R_{13}(z_1,z_3)R_{23}(z_2,z_3).
 \end{equation}
This is the spectral Yang-Baxter equation. The Bethe ansatz
solution exists, if the scattering matrix satisfies (\ref{40}), in
other words the matrix $\tilde{c}$ in (\ref{37}) and $\tilde{b}$
in (\ref{38}) are acceptable, only if the resulting $S$-matrices
satisfy (\ref{40}). This is a very restricted condition and needed
for having the solvability. The $S$-matrices (\ref{37}) and
(\ref{38}) are exactly the ones considered in \cite{13} and
\cite{14} respectively. Using the fact that $S^{(1)}$ is a
binomial of degree one with respect to $z_2$ and $S^{(2)}$ is a
binomial of degree one with respect to $z_1^{-1}=e^{-ip_1}$, it
can be shown that SYB equation (\ref{40}) for $S^{(1)}$ and
$S^{(2)}$ reduces to the non-spectral matrix equations
\be\label{43}
\tilde{c}_{12}[\tilde{c}_{12},\tilde{c}_{23}]=[\tilde{c}_{12},\tilde{c}_{23}]\tilde{c}_{23},
\end{equation}
and \be\label{44}
\tilde{b}_{23}[\tilde{b}_{23},\tilde{b}_{12}]=[\tilde{b}_{23},\tilde{b}_{12}]\tilde{b}_{12}.
\end{equation}
 which are the same as \cite{13} and \cite{14} respectively.
 The above equations are much simpler than (\ref{40}). So it is
 far simpler to seek the solutions of these equations than to seek
 those of (\ref{40}).

\subsection{Type 1 model}
The master equation (\ref{4}) with the boundary condition
\be\label{45} v_{\alpha_2} P_{\a_1\a_2}(x,x) =\sum_\b
c^{\b_1\b_2}_{\a_1\a_2}
 P_{\b_1\b_2}(x,x+1)
 \end{equation}
and relation (\ref{17}) describe consistently the following
reactions

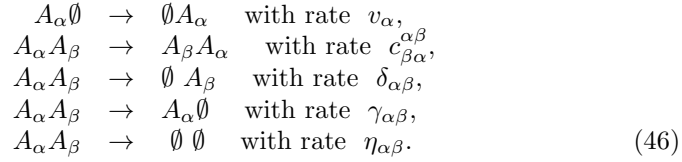
\bea\label{46}
 A_\alpha\emptyset &\rightarrow
&\emptyset A_\alpha \ \ \ \ {\rm with
 \ rate}\ \ v_{\a},\cr
  A_\alpha A_\beta &\rightarrow & A_\beta
A_\alpha \ \ \ \ {\rm with
 \ rate}\ \ c^{\alpha\beta}_{\beta\alpha},\cr
 A_\alpha A_\beta &\rightarrow &\emptyset \ A_\beta \
\ \ \ {\rm with
 \ rate}\ \ \delta_{\a\b},\cr
  A_\alpha A_\beta &\rightarrow &A_\alpha \emptyset  \ \ \ \ {\rm with
  \ rate} \ \ \gamma_{\alpha\beta},\cr
  A_\alpha A_\beta&\rightarrow &\ \emptyset\ \emptyset  \ \ \ \ {\rm with
  \ rate} \ \ \eta_{\alpha\beta}.
  \eea
A matrix $c$, or the above reactions, corresponds to an exactly
solvable exchange-reaction diffusion system on a one-dimensional
lattice, provided $c$ satisfies (\ref{17}) and $\tilde{c}$
satisfies (\ref{43}) and the non-diagonal elements are
nonnegative.
\subsection{Type 2 model}
The boundary condition

\be\label{47} v_{\alpha_2} P_{\a_1\a_2}(x,x) =\sum_\b
b^{\b_1\b_2}_{\a_1\a_2}
 P_{\b_1\b_2}(x-1,x)
 \end{equation}
and (\ref{18}) with the master equation (\ref{4}), introduce the
reactions as follows

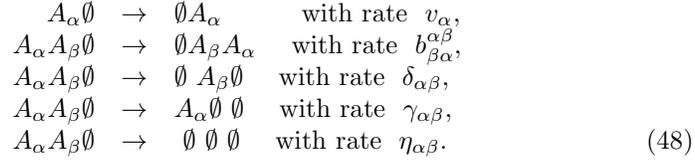
\bea\label{48}
 A_\alpha\emptyset &\rightarrow
&\emptyset A_\alpha \ \ \ \ \ \ \ \ \ {\rm with
 \ rate}\ \ v_{\a},\cr
  A_{\a}A_{\b}\emptyset &\rightarrow&
\emptyset A_{\b}A_{\a}  \ \ \ \ {\rm with
  \ rate} \ \ b^{\alpha\beta}_{\beta\alpha},\cr
A_\alpha A_\beta \emptyset &\rightarrow &\emptyset \ A_\beta
\emptyset \ \ \ \ {\rm with
 \ rate}\ \ \delta_{\a\b},\cr
  A_\alpha A_\beta \emptyset &\rightarrow &A_\alpha \emptyset \ \emptyset  \ \ \ \ {\rm with
  \ rate} \ \ \gamma_{\alpha\beta},\cr
  A_\alpha A_\beta\emptyset &\rightarrow &\ \emptyset\ \emptyset \ \emptyset  \ \ \ \ {\rm with
  \ rate} \ \ \eta_{\alpha\beta}.
 \eea
In brief, one can find the solvable systems corresponding to $b$
if the matrix $b$ satisfies a constraint (\ref{18}) and
$\tilde{b}$ satisfies (\ref{44}), and all the permissible elements
of $b$ (including diagonal) are nonnegative.

For more-than-two particle, now we consider a system consisting
$N$ particles of various species, with the evolution equation
\bea\label{49} \mathbf{\dot{P}}(x_1,\ldots,x_N) &=&
v_{\a_1}\mathbf{P}(x_1-1,\ldots,x_N)+\cdots+v_{\a_N}\mathbf{P}(x_1,\ldots,x_N-1)\cr
&-&(v_{\a_1}+\ldots+v_{\a_N})\mathbf{P}(x_1+\ldots+x_N)
 \eea
in the whole physical region, and boundary condition
\be\label{50} \mathbf{\dot{P}}(\ldots,x_k=x,x_{k+1}=x,\ldots)
=\tilde{b}_{k,k+1}\mathbf{P}(\ldots,x_k=x-1,x_{k+1}=x,\ldots),
 \end{equation}
 where $\tilde{b}_{k,k+1}$ is defined like (\ref{34}) and $\mathbf{P}$ is an
 $N$-tensor that the components of which are probabilities. It is
 seen that in this system, apart from the simple diffusion, there is a reaction
 between a block of $n+1$ adjacent particles
\be\label{51} A_{\a_0}\cdots A_{\a_n}\emptyset\longrightarrow
\emptyset A_{\g_0}\cdots A_{\g_n}\ \ \ \ \ {\rm with \ rate }\ \ \
\ \ (\tilde{b}_{n-1,n}\cdots
  \tilde{b}_{0,1})^{\a_0\cdots\a_n}_{\g_0\cdots\g_n},
\end{equation}
this comes from the fact that \bea\nonumber
&&\mathbf{P}(x_0=x,\ldots,x_{n-1}=x+n-1,x_n=x+n-1)\cr
&&=(\tilde{b}_{n-1,n}\cdots \tilde{b}_{0,1})\mathbf{P}
(x_0=x-1,\ldots,x_{n-1}=x+n-2,x_n=x+n-1). \eea
\be\label{52}\end{equation}
 Note the order of the matrices $\tilde{b}$. This order suggests that if a collection
of $n+1$ particles are adjacent, there is a probability that the
first particle pushes the second and changes the type of second
(and itself) and then it is the second (modified) particle that
interacts with the third.

\section{Two-particle system and exact solution} Let us focus on
the type 1 model. We consider $p=2$ case, but the argument can be
easily applied to arbitrary $p$. Taking $A_1=\mathcal{A}$ and
$A_2=\mathcal{B}$. One may label the two-particle states
$(\a_1,\a_2)$ as following
 \be\label{53}
 |1>=(1,1) \ , \ |2>=(1,2) \ , \ |3>=(2,1) \ , \ |4>=(2,2).
 \end{equation}
the most general matrix $c$ that satisfies (\ref{17}), can be
written as \be\label{54}
 c=\left(
 \begin{array}{cccc}
  v_{1}-\lambda_{1}& 0 & 0 & 0 \\
  0 &v_{1}-\lambda_{2}-c_{32}&c_{23}&0 \\
 0 & c_{32} &v_{2}-\lambda_{3}-c_{23}&0\\
 0 & 0 & 0 &v_{2}-\lambda_{4}
 \end{array}
 \right) \end{equation}
and therefore $\tilde{c}$ is (see (\ref{31}))\be\label{55}
 \tilde{c}=\left(
 \begin{array}{cccc}
  v_{1}-\lambda_{1}& 0 & 0 & 0 \\
  0 &v_{1}-\lambda_{2}-c_{32}&(\frac{v_1}{v_2})c_{23}&0 \\
 0 & (\frac{v_2}{v_1})c_{32} &v_{2}-\lambda_{3}-c_{23}&0\\
 0 & 0 & 0 &v_{2}-\lambda_{4}
 \end{array}
 \right) \end{equation}
where, according to (\ref{20}),
$\lambda_{i}=\delta_{i}+\gamma_{i}+\eta_{i}$.
 Now, we must check that under what conditions, $\tilde{c}$ satisfies the
(\ref{43}). If one writes (\ref{43}) as $RHS-LHS =0$, then one has
64 equations that must be solved for 16 variables. By using a
symbolic manipulator (e.g. MAPLE), one obtains the most general
solution. We express the class of parameters of this solution in
the following two sets as the given parameters (Gp) and the
arbitrary parameters (Ap)
 \bea\label{56}
&Gp&:\{c_{32}=0,
\delta_{1}=(\delta_{2}+\gamma_{2}+\eta_{2})-(\gamma_{1}+\eta_{1})\},\cr
&Ap&:\{v_1, v_2, c_{23}, (\gamma_{1}, \eta_{1}), (\delta_{2},
\gamma_{2}, \eta_{2}), (\delta_{3}, \gamma_{3}, \eta_{3}),
(\delta_{4}, \gamma_{4}, \eta_{4})\}.\ \eea Since all reaction
rates must be nonnegative, thus for $\delta_1\geq0$, one should
choose $(\lambda_2, \gamma_1, \eta_1)$ as $\lambda_2\geq
(\gamma_1+ \eta_1)$ . The interactions introduced by (\ref{54})
are
  \bea\label{57}
 \mathcal{A}\emptyset &\stackrel{v_1}\rightarrow & \emptyset \mathcal{A} \cr
 \mathcal{B}\emptyset  &\stackrel{v_2}\rightarrow &\emptyset \mathcal{B} \cr
 \mathcal{BA} &\stackrel{c_{23}}\rightarrow & \mathcal{AB} \cr
 \mathcal{AA} &\stackrel{\delta_1}\rightarrow & \emptyset \mathcal{A} \cr
 \mathcal{AA} &\stackrel{\gamma_1}\rightarrow &\mathcal{A} \emptyset  \cr
\mathcal{AA} &\stackrel{\eta_1}\rightarrow& \emptyset \
\emptyset\cr \mathcal{AB} &\stackrel{\delta_2}\rightarrow &
\emptyset \mathcal{B} \cr
 \mathcal{AB} &\stackrel{\gamma_2}\rightarrow &\mathcal{A} \emptyset  \cr
\mathcal{AB} &\stackrel{\eta_2}\rightarrow& \emptyset \
\emptyset\cr \mathcal{BA} &\stackrel{\delta_3}\rightarrow&
\emptyset\mathcal{A}\cr
 \mathcal{BA} &\stackrel{\gamma_3}\rightarrow &\mathcal{B} \emptyset  \cr
\mathcal{BA} &\stackrel{\eta_3}\rightarrow& \emptyset \
\emptyset\cr \mathcal{BB} &\stackrel{\delta_4}\rightarrow &
\emptyset \mathcal{B} \cr
 \mathcal{BB} &\stackrel{\gamma_4}\rightarrow &\mathcal{B} \emptyset  \cr
\mathcal{BB} &\stackrel{\eta_4}\rightarrow& \emptyset \ \emptyset
\eea
 The model built on the reactions (\ref{57}) is integrable. This is
a remarkable model because the rate of all the interactions is
arbitrary except, $\mathcal{AA}\rightarrow\emptyset\emptyset$,
that the rate of this process, $\delta_1$, depends on some other
processes.

{\bf Some subcases of this model:} In the above model, if we
abandon the annihilation processes ($\lambda_{i}=0$), we have the
first subcase as
 \bea\label{58}
 \mathcal{A}\emptyset &\stackrel{v_1}\rightarrow & \emptyset \mathcal{A} \cr
 \mathcal{B}\emptyset  &\stackrel{v_2}\rightarrow &\emptyset \mathcal{B} \cr
 \mathcal{BA} &\stackrel{c_{23}}\rightarrow & \mathcal{AB}
\eea this is the generalization of the reactions studied in
\cite{27} in which $c_{23}=v_2-v_1$. It has been showed in
\cite{26} that in the case of the single-species particles, each
particle has a distinct hopping rate according to its site instead
of its type (a particle in the position $x_i$ has a hopping rate
$v_i$). Now we consider a single-species system including the
annihilation processes, the boundary condition is
 \be\label{59}
v_{2} P(x,x) =(v_1-\lambda) P(x,x+1)\ \ \ \ \ \
\end{equation}
 with \be\label{60} \lambda=\delta+\gamma+\eta\
\end{equation}
 and the reactions are
 \bea\label{61}
\mathcal{A}\emptyset &\stackrel{v_1}\rightarrow & \emptyset
\mathcal{A} \cr \mathcal{AA} &\stackrel{\delta}\rightarrow &
\emptyset \mathcal{A} \cr
 \mathcal{AA} &\stackrel{\gamma}\rightarrow &\mathcal{A}
\emptyset\cr
 \mathcal{AA} &\stackrel{\eta}\rightarrow &
\emptyset \ \emptyset \eea
 the reactions (\ref{61}) make the second subcase. This subcase with $v_1=1$ and $\eta=0$ has been
considered in \cite{28}.

In the 2-species case for the type 2 model, the most general
matrix $b$ that satisfies (\ref{18}) and $\tilde{b}$ that is
constructed from matrix $b$, can be written as \be\label{62}
 b=\left(
 \begin{array}{cccc}
  v_{1}-\lambda_{1}& 0 & 0 & 0 \\
  0 &v_{1}-\lambda_{2}-b_{32}&b_{23}&0 \\
 0 & b_{32}&v_{2}-\lambda_{3}-b_{23}&0\\
 0 & 0 & 0 &v_{2}-\lambda_{4}
 \end{array}
 \right) \end{equation}
and\be\label{63}
 \tilde{b}=\left(
 \begin{array}{cccc}
  \frac{v_{1}-\lambda_{1}}{v_{1}^2}& 0 & 0 & 0 \\
  0 &\frac{v_{1}-\lambda_{2}-b_{32}}{v_{1}v_{2}}&\frac{b_{23}}{v_{2}^2}&0 \\
 0 & \frac{b_{32}}{v_{1}^2} &\frac{v_{2}-\lambda_{3}-b_{23}}{v_{1}v_{2}}&0\\
 0 & 0 & 0 &\frac{v_{2}-\lambda_{4}}{v_{2}^2}
 \end{array}
 \right) \end{equation}
 The matrix $\tilde{b}$ satisfies (\ref{44}) if we
 have two sets of Gp and Ap as
\bea\label{64} &Gp&:\{b_{23}=0,
\delta_{1}=\frac{v_{1}}{v_{2}}(\delta_{3}+\gamma_{3}+\eta_{3})-(\gamma_{1}+\eta_{1})\},\cr
&Ap&:\{v_1, v_2, b_{32}, (\gamma_{1}, \eta_{1}), (\delta_{2},
\gamma_{2}, \eta_{2}), (\delta_{3}, \gamma_{3}, \eta_{3}),
(\delta_{4}, \gamma_{4}, \eta_{4})\}.\ \eea This case is different
from that case we considered in the type 1 model. In the matrix
$c$ of the type 1 model the diagonal elements do not describe any
interactions but in this model all elements of matrix $b$ are
reaction rates. For example $c_{11} $ is not a reaction rate but
$b_{11}=v_1-\lambda_{1}$ is the reaction rate of
$\mathcal{A}\mathcal{A}\emptyset\rightarrow\emptyset\mathcal{A}\mathcal{A}$.
So one must choose the arbitrary parameters in such a manner that
all elements of $b$ (reaction rates) are nonnegative.

 \section{The conditional probability}
Assuming that the solvability condition (\ref{43}) or (\ref{44})
is satisfied. If at $t=0$ we have the initial configuration
$(y_1,\cdots,y_N)$, we require
\begin{equation}\label{65}
P(\vec{x};t=0)=\delta_{\vec{x},\vec{y}}
\end{equation}
and the probability $P(\vec{x};t)$ becomes the conditional
probability (the propagator) $U(\vec{x};t|\vec{y};0)$ and thus a
complete solution of the problem. It seems that the propagator is
\bea\label{66} U({\mathbf{x}};t|{\mathbf{y}};0)&=& \prod_{j=1}^N
v_{\a_j}^{x_j}e^{-v_{\a_j}t}
\int\frac{d^Np}{(2\pi)^N}f(\vec{p})e^{-E_{N}t}
 \sum_\sigma {\mathbf{A}}_\sigma e^{i\sigma
({\mathbf{p}}).{\mathbf{x}}} \cr &=&\prod_{j=1}^N
v_{\a_j}^{x_j-y_j}e^{-v_{\a_j}t}
\int\frac{d^Np}{(2\pi)^N}e^{-E_{N}t}e^{-i(\vec{p}.\vec{y})}
 \sum_\sigma {\mathbf{A}}_\sigma e^{i\sigma
({\mathbf{p}}).{\mathbf{x}}} \eea where the integration region for
each $p_i$ is $[0,2\pi]$ and $A_1=1$. In the above expansion
$f(\vec{p})$ is the coefficient of expansion which in the second
equality we choose it $\prod_{j=1}^N
v_{\a_j}^{-y_j}e^{-i(\vec{p}.\vec{y})}$, and by setting $p_j
\rightarrow p_j+i \varepsilon $, where one should consider the
limit $\varepsilon\rightarrow0^+$, (\ref{66}) reproduces the
required initial condition (\ref{65}). Using this propagator, one
can of course write the probability at the time $t$  in terms of
the initial value of probability
\begin{equation}\label{67}
|P({\mathbf{x}};t)\rangle=\sum_{y}U({\mathbf{x}};t|{\mathbf{y}};0)|P({\mathbf{y}};0)\rangle.
\end{equation}
Note that although $S^{(1)}$ and $S^{(2)}$ are similar to ones
considered in \cite{13,14}, but the propagator $U^{(1)} $and
$U^{(2)}$ are different since the energy spectrum and the
definition of the propagator of our models differ from those
considered there. For the two-particle sector, like reactions
(\ref{57}), there is only one matrix in the expression of
$U^{(i)}$s ($\tilde{c}$ in $U^{(1)}$ and $\tilde{b}$ in
$U^{(2)}$). So it can be treated as a $c$-number. Following
\cite{13,14} and \cite{17}, so by using the (\ref{66}), (\ref{36})
and (\ref{37}), the propagator for the two-particle system of the
type 1 model is

 \bea\label{68}
 U^{(1)}({\mathbf{x}};t|{\mathbf{y}};0)&=&\prod_{j=1}^2v_{j}^{x_j-y_j}e^{-v_{j}t}\int\frac{d^2p}{(2\pi)^2}e^{-E_{2}t-i(\vec{p}.\vec{y})}\cr
 &&\times\left[e^{i(p_1x_1+p_2x_2)}-\frac{1-e^{ip_2}\tilde{c}}{1-e^{ip_1}\tilde{c}}e^{i(p_2x_1+p_1x_2)}\right].
  \eea
where $E_2$ is obtained from (\ref{23}). Using the variable
$z_k=e^{ip_k}$, a simple contour integration yields \bea\label{69}
U^{(1)}({\mathbf{x}};t|{\mathbf{y}};0)&=&\prod_{j=1}^2v_{j}^{x_j-y_j}e^{-v_{j}t}\left\{\frac{t^{x_1-y_1}}{(x_1-y_1)!}\frac{t^{x_2-y_2}}
  {(x_2-y_2)!}\right. \cr
   &+& \left.\sum\limits_{l=1}^\infty\frac{t^{x_2-y_1+l}}{(x_2-y_1+l)!}\frac{t^{x_1-y_2}}
  {(x_1-y_2)!}\tilde{c}^{l}\left(-1+\frac{t\tilde{c}}{x_1-y_2+1}
  \right)\right\}
  \eea
and in the same way, for two-particle system of the type 2 model
one arrives at \bea\label{70}
U^{(2)}({\mathbf{x}};t|{\mathbf{y}};0)&=&\prod_{j=1}^2v_{j}^{x_j-y_j}e^{-v_{j}t}\left\{\frac{t^{x_1-y_1}}{(x_1-y_1)!}\frac{t^{x_2-y_2}}
  {(x_2-y_2)!}\right. \cr
   &+& \left.\sum\limits_{l=1}^\infty\frac{t^{x_2-y_1}}{(x_2-y_1)!}\frac{t^{x_1-y_2-l}}
  {(x_1-y_2-l)!}\tilde{b}^{l}\left(-1+\frac{(x_2-y_1)\tilde{b}}{t}
  \right)\right\}
\eea
 Now we concentrate in the type 1 model or $U^{(1)}$. As we said, the most general $\tilde{c}$ for
 the two-species system is the matrix (\ref{55}) with elements that determine by
 (\ref{56}). So we have
\be\label{71}
 \tilde{c}=\left(
 \begin{array}{cccc}
  v_{1}-\lambda_{2}& 0 & 0 & 0 \\
  0 &v_{1}-\lambda_{2}&(\frac{v_1}{v_2})c_{23}&0 \\
 0 & 0 &v_{2}-\lambda_{3}-c_{23}&0\\
 0 & 0 & 0 &v_{2}-\lambda_{4}
 \end{array}
 \right) \end{equation}
 Note that only the eigenvalues of $\tilde{c}$ with modulus
 $1$ cause singularities in $\mathbf{A_{\sigma}}$ at $p_j=0$ and
therefore in the integrand (\ref{66}). Corresponding to matrix
(\ref{71}), it is obvious that the eigenvalues of this matrix are
\bea\label{72} &&v_{1}-\lambda_{2},\cr&&v_{1}-\lambda_{2},\cr
&&v_{2}-\lambda_{3}-c_{23},\cr &&v_{2}-\lambda_{4}.
  \eea
Since all parameters that construct eigenvalues are arbitrary, so
the modulus of eigenvalues can be $1$ or different from $1$. To
investigate the large-time behavior of the propagator $U^{(1)}$,
it is useful to decompose the vector space on which $\tilde{c}$
acts, into two subspaces invariant under the action of
$\tilde{c}$. The first subspace corresponding to eigenvalues with
modulus $1$, and the second invariant subspace corresponding to
eigenvalues with modulus different from $1$. This decomposition
can be done by introducing two projections $Q$ and $R$, satisfying
 \bea\label{73}
 Q+R=1,\cr
 QR=RQ=0,\cr
 [\tilde{c},Q]=[\tilde{c},R]=0.
 \eea
 $Q$ projects on the first subspace and $R$ projects on the second.
 Following \cite{13}, we multiply $U^{(1)}$ by $Q+R=1$:
 \bea\label{74}
U^{(1)}(\mathbf{x};t|\mathbf{y};0)=U^{(1)}Q+U^{(1)}R.
  \eea
In the term multiplied by $R$, one can treat $\tilde{c}$ as a
number with modulus different from $1$, Thus in this term, there
is no pole in $S^{(1)}$ and hence in $\mathbf{A_{\sigma}}$. So the
integrand in (\ref{66}) is nonsingular at $p_{j}=0$, which have
the main contributions at large times. Setting $p_j=0$ in
$S^{(1)}$ as an approximation to arrive at

\begin{equation}\label{75}
S^{(1)}\approx-1,
\end{equation}
and
\begin{equation}\label{76}
A_{\sigma}\approx(-1)^{[\sigma]}.
\end{equation}
One can also approximate $E$ as
\begin{equation}\label{77}
E(\mathbf{p})\approx\sum_{j=1}^{2}\left(-1+ip_j+\frac{p_{j}^{2}}{2}\right).
\end{equation}
So, the second term of (\ref{74}) for large times results in
 \bea\label{78}
U^{(1)}R&=&\prod_{j=1}^2v_{j}^{x_j-y_j}e^{-v_{j}t}\frac{1}{2\pi
 t}\left\{e^{-[(x_{1}-y_{1}-t)^{2}+(x_{2}-y_{2}-t)^2]/(2t)}\right.\cr
  &&\left.-e^{-[(x_{1}-y_{2}-t)^{2}+(x_{2}-y_{1}-t)^2]/(2t)}\right\}R
  \ \ \ t\longrightarrow\infty.
\eea which is independent of $\tilde{c}$. So at large times, the
second term tends to zero faster than $t^{-1}$ and only the first
term of (\ref{74}) survives.

If one chooses the class of parameters in such a manner that the
only eigenvalue of $\tilde{c}$ with modulus $1$ is $1$, then
$U^{(1)}$ has a simple behavior for $t\rightarrow\infty$. In this
case, $\tilde{c}Q=Q$, and one can simplify $U^{(1)}$ to
find\bea\label{79}
 U^{(1)}({\mathbf{x}};t|{\mathbf{y}};0)&=&\prod_{j=1}^2v_{j}^{x_j-y_j}e^{-v_{j}t}\left\{\frac{t^{x_1-y_1}}{(x_1-y_1)!}\frac{t^{x_2-y_2}}
  {(x_2-y_2)!}\right. \cr
   &+& \left.\sum\limits_{l=1}^\infty\frac{t^{x_2-y_1+l}}{(x_2-y_1+l)!}\frac{t^{x_1-y_2}}
  {(x_1-y_2)!}\left(-1+\frac{t}{x_1-y_2+1}
  \right)\right\}Q \cr
  &&t\longrightarrow\infty.
    \eea
This is the propagator corresponding to a single-species
asymmetric simple exclusion process with PDHR multiply by $Q$. In
fact (\ref{79}) is the case of the model in \cite{26}, times Q.
\section{Conclusion}
We have defined two generalized totally asymmetric exclusion
processes, in which we have $p$-species particles that interact
and diffuse where each particle $A_{\a}$ has its own intrinsic
hopping rates $v_{\a}$. We have introduced two types of boundary
conditions in terms of two $p^2 \times p^2$ matrices $c$ and $b$
respectively, that led to the two new models, i.e.
reaction-diffusion and drop-push systems with PDHR. At first we
have assumed that these interactions preserve the total number of
particles, so we have obtained the constraint $\sum_{\vec{\a}}
 F^{\b_1\b_2}_{\a_1\a_2} = v_{\b_{1}}$ ( here and elsewhere $F$ denotes either $b$ or
 $c$). Next we have shown that when one violates the conservation of
 particles, the mentioned constraint changes into $ \sum_{\vec{\a}}F^{\b_1\b_2}_{\a_1\a_2} =
 v_{\b_{1}}-\lambda_{\b_{1}\b_{2}}$ and the annihilation processes
 add to the previous reactions of new models, where
 $\lambda_{\b_{1}\b_{2}}$ is the sum of all annihilation
 processes. We have also shown that only for the states $\vec{\b}=\vec{\a}$ and
 $\vec{\b}=(\a_{2},\a_{1})$, $F$ introduces the new model with
 PDHR and has non-zero elements. These states resulted in
  that matrix $\tilde{F}$, a version of $F$ that constructs $S$-matrix and determines the
 coefficient ${\mathbf{A}}_\sigma$, must satisfy the non-spectral equation matrix for solvability of our models in the sense of the Bethe ansatz.
 Then we have considered the two-particle systems and
 obtained the permissable reaction rates by checking the solution
 of the non-spectral matrix equation. Finally the
 conditional probability of our new models has been obtained and for the two-particle sector, the exact expression and
 specially its large-time behavior have been calculated. We have shown that the propagator of
 2-species reaction-diffusion system with PDHR
 in large times is equivalent to the propagator of the single-species
asymmetric simple exclusion process with PDHR.


\begin{thebibliography}{9}
\bibitem {1} B. Derrida, S. A. Janowsky, J. L. Lebowitz and E. R. Speer, 1993 \textsl{Europhys. Lett}. {\bf 22} 651
\bibitem {2} P. A. Ferrari and L. R. G. Fontes, 1994 \textsl{Probability Theory Related Fields.} {\bf 99} 305
\bibitem {3} J. M. Burgers, 1974 {\it The Nonlinear Diffusion Equation }
(Boston: D. Reidel)
\bibitem {4} J. Krug and H. Spohn, 1991 {\it in Solids Far From Equilibrium }, edited by C. Godreche (Cambridge: Cambridge University
Press).
\bibitem {5} K. Nagel, 1996 \textsl{Phys. Rev.} E {\bf 53} 4655
\bibitem {6} C. T. MacDonald, J. H. Gibbs and A. C. Pipkin, 1968 \textsl{Biopolymers.} {\bf 6} 1
\bibitem {7} G. M. Sch\"{u}tz, 1997 \textsl{J. Stat. Phys.} {\bf 88} 427
\bibitem {8} M. Alimohammadi, V. Karimipour and M. Khorrami, 1998 \textsl{Phys. Rev.} E {\bf 57} 6370
\bibitem {9} T. Sasamoto and M. Wadati, 1998  \textsl{Phys. Rev.} E {\bf 58}, 4181
\bibitem {10} T. Sasamoto and M. Wadati, 1998 \textsl{J. Phys. }A {\bf 31}, 6057
\bibitem {11} A. M. Povolotsky, V. B. Priezzhev and C.-K. Hu, 2003 \textsl{J.
Stat. Phys.} {\bf 111} 1149
\bibitem {12} M. Alimohammadi and N. Ahmadi, 2000 \textsl{Phys. Rev.} E {\bf 62} 1674
\bibitem {13} F. Roshani and M. Khorrami, 2001 \textsl{Phys. Rev.} E {\bf 64} 011101
\bibitem {14} F. Roshani and M. Khorrami, 2003 \textsl{Eur. Phys. J.} B {\bf 36} 99
\bibitem {15} M. Alimohammadi, 2004 \textsl{Eur. Phys. J.} B {\bf 42} 415
\bibitem {16} M. Alimohammadi, V. Karimipour and M. Khorrami, 1999 \textsl{J. Stat. Phys.} {\bf 97} 373
\bibitem {17} M. Alimohammadi and Y. Naimi, 2005 \textsl{J. Math. Phys.} {\bf 46} 053306
\bibitem {18} I. Benjamini, PA. Ferrari and C. Landim, 1996 \textsl{Stoc. Proc. Appl.} {\bf
61} 181
\bibitem {19} M. R. Evans and T. Hanney, 2005 \textsl{J. Phys. A:
Math. Gen.} {\bf 38} R195
\bibitem {20} S. Grosskinsky, G. M. Sch\"{u}tz and H. Spohn, 2003 \textsl{J. Stat. Phys.} {\bf 113} 389
\bibitem {21} C. Godr\`{e}che, 2003 \textsl{J. Phys. A: Math. Gen.} {\bf 36} 6313
\bibitem {22} E. Levine, D. Mukamel and G. M. Sch\"{u}tz, 2005 \textsl{J. Stat. Phys.} {\bf
120} 759
\bibitem {23} J. Kaupuzs, R. Mahnke and R. J. Harris, 2005 \textsl{Phys. Rev.} E {\bf72} 056125
\bibitem {24} R. J. Harris, A. R\'{a}kos and G. M. Sch\"{u}tz, 2005 \textsl{J. Stat. Mech.} P08003
\bibitem {25} C. Godr\`{e}che and J. M. Luck, 2005 \textsl{J. Phys.} A {\bf 38} 7215
\bibitem {26} A. R\'{a}kos and G. M. Sch\"{u}tz, 2006 \textsl{Markov Processes and Related
Fields.} {\bf 12} 323
\bibitem {27} V. Karimipour, 1999 \textsl{Europhys. Lett.} {\bf 47} 501
\bibitem {28}  F. Roshani and M. Khorrami, 2002 \textsl{J. Math. Phys.} {\bf43}
2627

\end{thebibliography}
\end{document}